\documentclass{vgtc}                          

\ifpdf
  \pdfoutput=1\relax                   
  \usepackage{graphicx}                
  \DeclareGraphicsExtensions{.pdf,.png,.jpg,.jpeg} 
\else
  \usepackage{graphicx}                
  \DeclareGraphicsExtensions{.eps}     
\fi%

\graphicspath{{figures/}{pictures/}{images/}{./}} 

\usepackage{microtype}                 
\PassOptionsToPackage{warn}{textcomp}  
\usepackage{textcomp}                  
\usepackage{mathptmx}                  
\usepackage{times}                     
\usepackage{cite}                      
\usepackage{tabu}                      
\usepackage{booktabs}                  

\usepackage{xcolor,colortbl}
\usepackage[draft]{todonotes}   
\colorlet{red}{black}

\onlineid{0}

\vgtccategory{Research}

\vgtcinsertpkg

\preprinttext{Online Submission ID: 1237}

\title{FeatureExplorer: Interactive Feature Selection and Exploration\\ of Regression Models for Hyperspectral Images}

\author{Jieqiong Zhao\thanks{e-mail: zhao413@purdue.edu}\\ %
        \scriptsize Purdue University %
\and Morteza Karimzadeh\thanks{e-mail: karimzadeh@colorado.edu}\\ %
     \scriptsize University of Colorado Boulder %
\and Ali Masjedi\thanks{e-mail: amasjedi@purdue.edu}\\ %
     \scriptsize Purdue University
\and Taojun Wang\thanks{wang3241@purdue.edu}\\
     \scriptsize Purdue University\\
\and Xiwen Zhang\thanks{zhan2977@purdue.edu}\\
     \scriptsize Purdue University
\and Melba M. Crawford\thanks{e-mail: mcrawford@purdue.edu}\\
     \scriptsize Purdue University
\and David S. Ebert\thanks{e-mail: ebertd@purdue.edu}\\
     \scriptsize Purdue University}

\abstract{Feature selection is used in machine learning to improve predictions, decrease computation time, reduce noise, and tune models based on limited sample data. 
In this article, we present FeatureExplorer, a visual analytics system that supports the dynamic evaluation of regression models and importance of feature subsets through the interactive selection of features  in  high-dimensional feature spaces typical of hyperspectral images.
The interactive system allows users to iteratively refine and diagnose the model by selecting features based on their domain knowledge, interchangeable (correlated) features, feature importance, and the resulting model performance.
} 

\begin{document}

\firstsection{Introduction}
\maketitle

Machine learning methods are increasingly used to analyze big data. However, many of these methods are used as black boxes (primarily because of the way current computational libraries present the models/results). Therefore, domain users may not understand how the results are generated, and may not trust the models.
These problems are further complicated by insufficient data samples and the curse of dimensionality.
Feature selection is often adopted to improve these models by identifying the relevant features that contribute the most to the prediction results while removing noisy, irrelevant, and less important features.

In this paper, we present FeatureExplorer, a visual analytics system to support interactive feature selection and model evaluation for remotely-sensed data. 
To design this system, we collaborated with remote sensing experts and plant scientists whose goal was to predict plants' wet biomass using data recorded in hyperspectral imagery. These domain experts needed to identify the predictive ability and interchangeability of key features derived from hyperspectral images (and their underlying wavelengths) for biomass prediction. It was challenging to investigate such high dimensional datasets and regression models without visual analytics tools, which motivated the design of FeatureExplorer. It enables experts to trace the regression models back to the key contributing features (hyperspectral indices), and ultimately the pertinent image wavelengths (among a large number of bands), along with options for interactive manipulation, feature selection, and model evaluation based on domain knowledge. 

Our system supports integrated visual exploration and selection of features through the analysis of: (1) linear relationships  among features using a correlation matrix; (2) distribution of any pair of two features using a scatterplot enhanced with Kernel Density Estimation (KDE) visualizations; (3) feature importance ranking for non-linear relationships based on a combination of a feature selection method (Recursive Feature Elimination (RFE)) and a regression model (Support Vector Regression (SVR)).


We summarize the contributions of this paper as follows:
\begin{itemize}
\item An interactive system supporting dynamic feature exploration and selection based on univariate and multivariate feature analysis with integrated regression models, reducing the large number of features to a few key ones that can be used for improved modeling and future data collection and analysis.
\item \textcolor{black}{Experimental results comparing various machine learning methods for predicting biomass using hyperspectral indices.}
\item A workflow for identifying key hyperspectral indices and the original reflectance values used in index calculations.
\item \textcolor{black}{A case study of the use of the platform by domain experts for hyperspectral image analysis to predict plant wet biomass.}

\end{itemize}

\section{Background}
Biomass is an important plant characteristic that helps with crop monitoring, yield estimation, and indicating plant growing conditions, and is quantified based on the above-ground weight of a plant before dehydration.
In the case of sorghum (the main plant in our study), wet biomass  determines the amount of ethanol product. To identify superior plant varieties for breeding, biomass can be manually measured at the end of the growing season; however, this traditional method is time consuming, expensive, and retrospective. Instead, hyperspectral images collected by Unmanned Aerial Vehicles (UAVs) throughout the season can be used to predict the final biomass.
Remote sensing experts in our team collected high resolution hyperspectral images acquired multiple times (from June to Sept.) over 14 acres of experimental sorghum fields with 830 varieties in the 2017 growing season.
The ground truth wet biomass was measured once at the end of the growing season (Oct. $15^{th}$).

\textcolor{red}{A hyperspectral image captures a spectrum that covers wavelengths ranging from 400 nm to 1000 nm in 2.2 nm increments for each pixel (272 bands). 
The original collected 272 bands are continuous narrow bands, which are highly correlated with neighboring ones.
To reduce the dependency among these bands, we adopted hyperspectral indices based on domain practice.}
Specifically, we utilize the 36 hyperspectral vegetation indices listed in ~\cite{liang2015estimation}.
\textcolor{red}{Each index is derived from several bands values and based on a unique plant biophysical meaning.
However, some indices have closely-related calculation formulas.
} More information about the sensors, data pre-processing, and feature extraction is available in~\cite{masjedi2018sorghum, elbahnasawy2018multi, zhang2017prediction}.
\section{Related Work}
Feature selection methods can be generally divided into four categories: filter methods, wrapper methods, embedded methods, and hybrid methods~\cite{Chandrashekar2014}.
The filter and wrapper categories are relevant to our work; therefore, we will focus on them here.
Pearson's correlation coefficient is a popular filtering method for narrowing down features to the ones with high (linear) correlation with the dependent variable. However, correlated but redundant features may be selected, and the coefficient is unable to characterize nonlinear relationships.
Wrapper methods use regression or classification models to find an optimal feature subset by iteratively adding or removing features. 
The combination of learning models (e.g. SVR) and wrapper methods (e.g. RFE) has traditionally been  used for automatic feature selection ~\cite{duan2005multiple, Ding:2011:SBF:1943363.1943368}.

Several visualizations have been proposed for feature selection, including correlation matrices~\cite{Friendly2002}, feature clustering~\cite{Yang2003}, feature ranking~\cite{JinwookSeo2004, Piringer2008, Johansson2009}, scatterplot matrices~\cite{Elmqvist2008}, and dimensionality reduction~\cite{Brushing11}.
A few visual analytics systems have leveraged a combination of automatic and visual feature selection techniques.
RegressionExplorer~\cite{dingen2019regressionexplorer} is one such system for inspecting logistic regression models.
Other systems have been proposed to support exploring linear relationships among features ~\cite{Guo2009,Piringer2010,Barlowe2008}. BEAMES~\cite{dasbeames} is another multi-model system that enables users to interactively compare different types of models with various hyper-parameters (e.g., logistic regression vs. Bayesian regression models), while allowing users to interactively weigh data instances and features.
\textcolor{red}{INFUSE~\cite{Krause2014} enables the ensemble of multiple feature selection methods by visualizing features importance as determined by various feature selection methods in a radial glyph. Our focus, however, is to support domain experts in efficiently reducing a high-dimensional feature space into key feature subsets, and tracing back the features to the underlying wavelengths for incorporating domain knowledge.}

Partition-based visual analytics systems~\cite{May2011, Muhlbacher2013} primarily focus on the interactive exploration of local structures and relationships of independent and target variables, appropriate for lower feature space dimensions. They are aimed at closer inspection of limited numbers of selected features for optimal distribution partitioning and model building. 
However, our focus is on high dimensions (of both data instances and feature space). \textcolor{red}{
Our system's integrated hierarchical clustering and matrix visualizations facilitate the quick identification of (a) influential feature subsets (either already selected or missing) for model building, (b) the interchangeable features within those subsets, and (c) detailed feature distribution and importance.}

\section{Design Goals}
\textcolor{red}{We collaborated with three remote sensing experts: two Ph.D. students and a senior faculty member with expertise in hyperspectral image analysis for agronomy.
Traditionally, they predict biomass using automated feature selection algorithms and regression models.
Oftentimes, optimally tuning these  algorithms requires large numbers of data samples, which are expensive to collect. It is challenging to build a model that performs well for all kinds of hybrid varieties, plants in different locations, or at different growing stages/conditions with limited samples.
Therefore, the domain experts needed to identify the key hyperspectral features to achieve stable, credible, and accurate prediction results, using both automated methods and 
their domain knowledge to inspect the relationship among features, the importance of features, and  trace the hyperspectral indices back to the biophysical space.
Hyperspectral indices indicate meaningful chemical concentrations in plants, which can be applied to differentiate plant varieties.}
\textcolor{red}{
The domain experts also expressed the need for clustering features, dynamic feature selection, and model performance comparisons with and without feature selection.
We derived the following design goals to fulfill these requirements:}



\begin{enumerate}
\item [\textbf{DG1}] Interactive exploration of features, including feature density distributions and relationships among multivariate features.
\item [\textbf{DG2}] Identification of important features such as influential hyperspectral indices and the underlying wavelengths that contribute to the prediction of wet biomass.
\item [\textbf{DG3}] Direct manipulation and refinement on subset of features through interactively adding and removing specific features.
\item [\textbf{DG4}] Evaluation of regression results with ground truth for subset of selected features versus full set of feature.
\end{enumerate}

\textcolor{red}{These requirements were formalized into design mock-ups using visualizations already familiar to domain users based on their request. We then implemented the design, and made minor modifications according to feedback from domain experts, as described below.}



\begin{figure}[bht]
  \centering
  \resizebox{0.8\columnwidth}{!}{\includegraphics{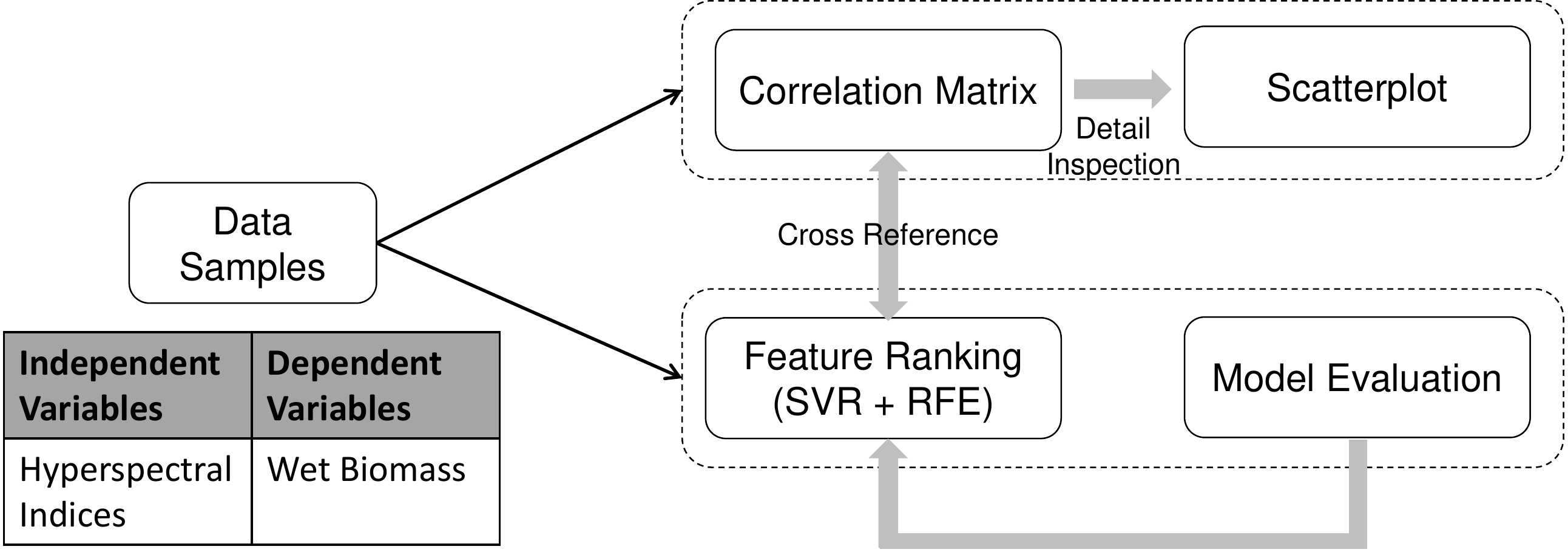}}
  \caption{The components diagram of FeatureExplorer.}
  \label{fig:workflow}
  \vspace{-1.5em}
\end{figure}

\begin{figure*}[thb]
  \centering
  \resizebox{\textwidth}{!}{\includegraphics{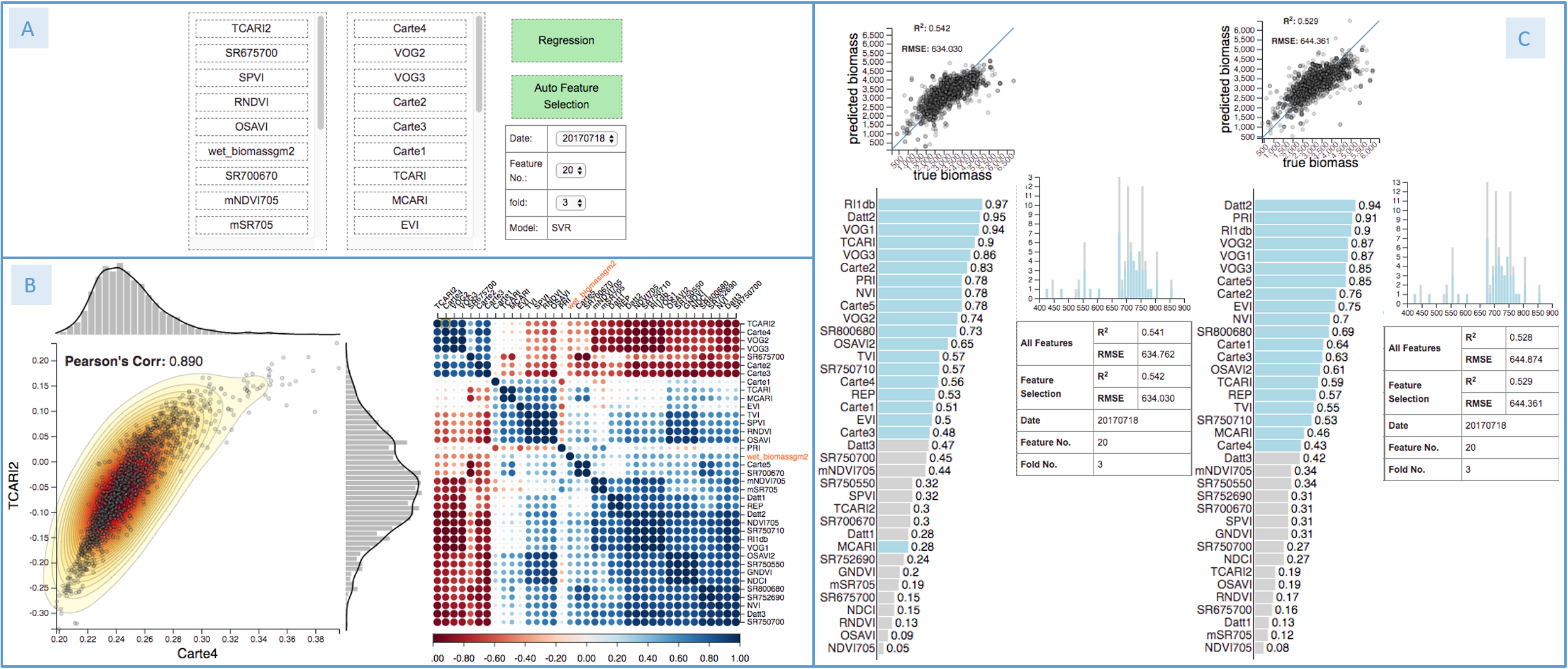}}
  \caption{FeatureExplorer overview: (A) the control panel with a list of unselected features, a list of selected features, a regression button, an automatic feature selection button; (B) feature correlation panel with a correlation matrix and a scatterplot; (C) evaluation panel with a scatterplot of ground truth and predicted values, a horizontal bar chart showing the importance score of each feature, a histogram showing the frequency of used pertinent wavelengths, a table displaying the results with and without feature selection.}
  \label{fig:interface}
  \vspace{-1.5em}
\end{figure*}

\section{FeatureExplorer}
In this section, we first explain how our system addresses the design goals, and then elaborate on the frontend user interface and backend analytics components of FeatureExplorer.

\subsection{\textcolor{red}{Workflow}}
Figure 1 presents the system components in FeatureExplorer, and our process.
As shown in Figure~\ref{fig:workflow}, FeatureExplorer supports the analysis of both linear and non-linear relationships (DG1, DG2). 
To visualize feature relationships, a correlation matrix serves as an overview to render the Pearson's correlation coefficient for all pairs of features.
Users can click on any cell for a detailed inspection of any particular pair of features. 
For non-linear relationship analysis, Support Vector Regression and Recursive Feature Elimination (SVR + RFE) provide feature importance ranking.
Users can compare and analyze the ranking results and use the synthesized information to add or remove features (DG3). 
$R^2$ and Root Mean Square Error (RMSE) are calculated to show the regression models' performance with the selected subset of features (DG4).
\textcolor{red}{After initial implementation, users requested the capability to adjust the number of folds in cross validation, to compare the performance of regression models with  a selected subset of features versus with all features, and to map hyperspectral indices to original wavelengths.}
This way, users can utilize the gained insights from the interactive exploration process to identify the underlying pertinent wavelengths, and strategize ways to collect only pertinent data in the future to save cost and time.


\begin{figure*}[thb]
    \centering
    \resizebox{0.85\textwidth}{!}{\includegraphics{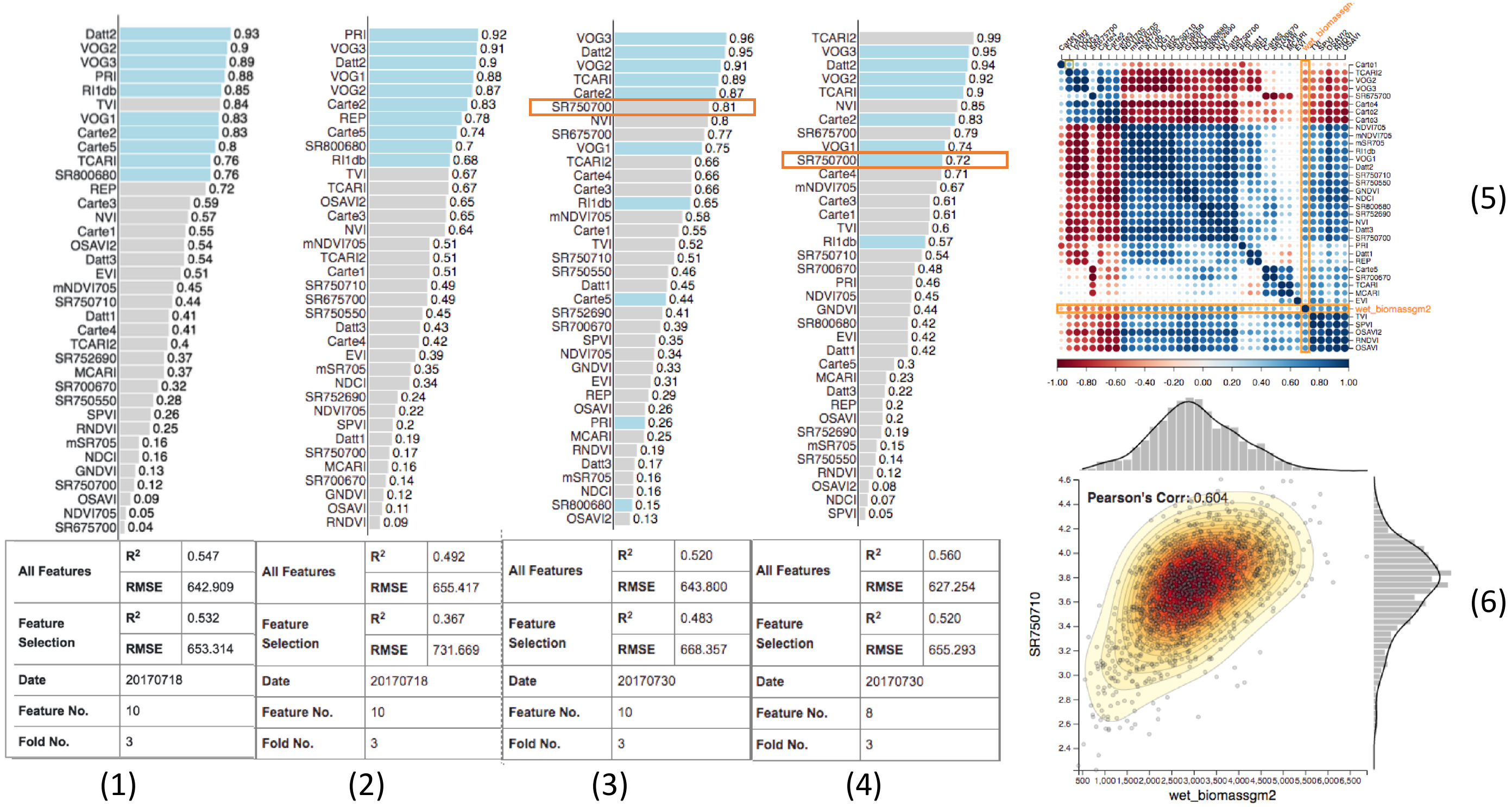}}
    \caption{Case study in using FeatureExplorer for two hyperspectral datasets.}
    \label{fig:case_study}
    \vspace{-1.5em}
\end{figure*}

\subsection{User Interface}
Figure~\ref{fig:interface} illustrates the user interface that 
contains three panels: (A) a control panel, (B) a correlation panel, and (C) an evaluation panel. 
As we described in the previous section, the two latter panels are separated based on 
the linearity of the relationship between input features and predicted variables.
\textcolor{black}{In this section, we describe the views individually, and will showcase the integrated use of these views in a use case in Section~\ref{sec:case_study}.}

In the correlation panel, a correlation matrix shows the Pearson's correlation coefficient between any pair of features.
The coefficient value is double-encoded using two visual channels (color and radius) for better usability.
Hierarchical clustering groups the features based on the similarity of correlations to other features. This helps users identify representative pairs from each cluster while minimizing the chances of including other similarly correlated pairs. 
While providing a good overview, a single correlation value does not provide sufficient information for interpreting the relationship between two features. To address this, users can click on any cell to see the scatterplot of the selected two features.
The system uses both histograms and KDE to illustrate the marginal distribution of univariate features at the edge of the histogram.
We also overlaid a 2D KDE on the scatterplot to better visualize the distribution of two features.
\textcolor{red}{The marginal distributions and KDE contours are beneficial in understanding general data patterns. 
The domain users pointed out that exploring the hyperspectral index vs. wet biomass scatterplot  could help them investigate whether the index captures the variation across high and low biomass values.
}

At the top part of the evaluation panel, a scatterplot shows ground truth values against predicted results along with $R^2$ and RMSE values.
\textcolor{red}{With this graph, domain users identified that the regression model does not perform well on extremely high or low biomass values.}
The horizontal bar graphs show the feature importance score for each input feature (using SVR + RFE), and the light blue rectangles indicate selected features. 
The histogram beside the bar graphs shows the frequency of using pertinent reflectance (raw data) to derive the indices in the subset of selected features over the wavelength range of 400 nm to 900 nm.
This enables domain experts to trace back the selected features to the wavelengths that are utilized to derive the indices.
Moreover, a table shows performance comparison for a subset of selected features versus all features based on the same data partition (training vs. testing) and regression model.

As we mentioned before, the correlation matrix and the SVR + RFE bar graphs provide different rankings, the former for linear relationships and the latter for non-linear models. 
Users can refer to both to adjust the subset of selected features.
In the control panel, the leftmost list shows unused features, and the list in the middle shows the selected ones. 
Users can drag and drop features between these two lists and evaluate the results on the fly.
To avoid exhaustive feature searching at the beginning by the users, the system enables an initial automatic feature selection method based on SVR + RFE. 


\begin{table}[htb]
\small
\begin{tabular}{c|c|c|c|c|c|c}
\hline
\textbf{Date} & 
\textbf{Ridge} & 
\begin{tabular}[c]{@{}c@{}}\textbf{Elastic} \\ \textbf{Net}\end{tabular} & 
\begin{tabular}[c]{@{}c@{}}\textbf{Partial}\\ \textbf{Least}\\ \textbf{Square}\end{tabular} & 
\textbf{SVR}  & 
\begin{tabular}[c]{@{}c@{}}\textbf{Random}\\ \textbf{Forest}\end{tabular} & 
\textbf{AdaBoost} \\ 
\hline
06/21 & \textbf{0.20}                                        & 0.13                                                  & \textbf{0.20}                                                    & \textbf{0.20} & \textbf{0.20}                                           & 0.15     \\ 
06/27 & \textbf{0.25}                                        & 0.16                                                  & \textbf{0.25}                                                    & 0.24          & 0.23                                                    & 0.18     \\ 
07/04 & \textbf{0.27}                                        & 0.17                                                  & \textbf{0.27}                                                    & \textbf{0.27} & 0.26                                                    & 0.19     \\ 
07/18 & 0.51                                                 & 0.23                                                  & 0.51                                                             & \textbf{0.53} & 0.44                                                    & 0.36     \\ 
07/30 & 0.51                                                 & 0.28                                                  & 0.52                                                             & \textbf{0.55} & 0.49                                                    & 0.45     \\ 
08/08 & 0.53                                                 & 0.34                                                  & 0.53                                                             & \textbf{0.56} & 0.50                                                    & 0.45     \\ 
08/14 & 0.53                                                 & 0.35                                                  & 0.53                                                             & \textbf{0.54} & 0.51                                                    & 0.45     \\ 
08/23 & \textbf{0.54}                                        & 0.34                                                  & \textbf{0.54}                                                    & \textbf{0.54} & \textbf{0.54}                                           & 0.50     \\ 
09/10 & \textbf{0.52}                                        & 0.32                                                  & \textbf{0.52}                                                    & \textbf{0.52} & \textbf{0.52}                                           & 0.47     \\ 
09/24 & 0.51                                                 & 0.35                                                  & 0.51                                                             & \textbf{0.52} & 0.51                                                    & 0.45     \\ \hline
\end{tabular}
\caption{Comparison of average $R^2$ for 100 trials among multiple regression models on 10 datesets.}
\vspace{-3em}
\label{tab:model_comparison}
\end{table}

\subsection{Regression Models}
After testing several regression models including Ridge, Elastic Net, Partial Least Squares, SVR, Random Forest, and AdaBoost, we found that SVR~\cite{scholkopf2002learning} outperforms other models for predicting biomass from hyperspectral indices for most dates. \textcolor{black}{The results of $R^2$ for these regression models are listed in Table~\ref{tab:model_comparison}. Since $R^2$ and RMSE are highly correlated (higher $R^2$ means lower RMSE), we only report the $R^2$.
Based on the results, we decided to integrate SVR + RFE (for automatic feature selection) into the system.}

The system runs k-fold cross validation for model evaluation.
For each training of the SVR model, the system first runs a grid search with a Radial Basis Function (RBF)~\cite{alpaydin2009introduction} kernel
to select the best model hyperparameters that maximize $R^2$, and then performs initial feature selection on that model~\cite{liu2011feature}.
The RFE ranks the features based on their contributions in the regression model, and the system transforms these ranks to scores in the range of [0, 1], 0 meaning no contribution and 1 meaning the most important feature in the model. 

\textcolor{black}{We use  Equation~\ref{equ:feature_ranking} to compute the ranking score of a feature, where k is the number of folds, d is the number of dimensions in the feature space, and r denotes the ranking determined by RFE.
The RFE method outputs the ranking of features in a sequential order from the most important to least; the most important feature has a ranking of 1 and the least important feature has a ranking of d.
The numerator of Equation~\ref{equ:feature_ranking} sums the normalized ranking (mapping values in [1, d] to [0, 1])}, which is then divided by k to calculate the average of these scores for multiple runs (in cross-fold validation). We use this RankingScore in feature importance visualization (the horizontal bar graphs).
\begin{equation}
Ranking Score = \frac{\sum_{i=1}^{k}\frac{(d + 1 - r_i) - 1}{d - 1}}{k}
\label{equ:feature_ranking}
\end{equation}


\section{Case Study}\label{sec:case_study}
A remote sensing expert in our team used FeatureExplorer to investigate hyperspectral indices for biomass prediction.
He aimed to determine which indices were the most predictive ones, and if he could reduce a combination of 36 features down to 10 key features while understanding their biophysical meanings in collaboration with a plant scientist.
He used 10 hyperspectral images collected from June 21st to Sept. 24th in 2017 to investigate whether the important subset of hyperspectral indices changed in each image set.
First, he started with one dataset (July 18th) and applied automatic feature selection for 20 features (out of 36 total), and found that performance using 20 features was slightly better than when using all 36 features.
Then, he applied automatic feature selection, limiting to 3 features. The regression performance ($R^2$) dropped significantly (higher RMSE).
Based on ranked feature sets and the correlation matrix, he added 4 features that had high importance scores and low correlation among them.
\textcolor{black}{These 4 features were selected from different clusters in the correlation matrix, since he wanted the regression model to learn useful information from diverse features.}
The performance of the model improved.
After adding up to 10 features, the performance of the regression model was almost equivalent to its performance when using 20 features (Figure~\ref{fig:case_study}(1)).
He then tested whether applying automatic selection limited to 10 features would lead to similar results; it turned out that the manually selected features outperformed
the automatic selection (Figure~\ref{fig:case_study}(2)).

Next, he applied the same subset of features on another hyperspectral image (July 30th) that was captured 12 days after the first one.
He found that wet biomass had stronger correlations with most hyperspectral indices (the correlation matrix shown in Figure~\ref{fig:case_study}(5)) compared with the first dataset (the correlation matrix shown in Figure~\ref{fig:interface}(B)).
The regression model performed better on the second dataset than the first one because the plants were at a different growing stage~\cite{gerik2003sorghum} and their reflectance had changed~\cite{brandao2015spectral}.
Tuning the regression model on the second dataset with the 10 features selected during analyzing the first dataset did not improve the prediction results; however, the performance of the regression model did not drop dramatically (Figure~\ref{fig:case_study}(3)).
By carefully examining the correlation matrix for the second dataset, he found 3 features that did not have high correlations with biomass.
After removing these 3 features and adding another feature which had a high importance score and high correlation with biomass, the model's performance improved significantly (Figure~\ref{fig:case_study}(4)). This indicates the human-in-the-loop can improve the predictive performance of the regression model.


\section{Conclusion and Future Work}
We presented a visual analytics system for the exploration, ranking, and selection of features in integrated regression models supporting analysis on linear and non-linear relationships. The system provides initial automated feature selection, and enables users to dynamically change, compare and evaluate models' performance based on user-specified subsets of features. We demonstrated the successful use of the system by remote sensing experts to identify important hyperspectral indices at various plant growth stages for predicting the biomass at the end of the growing season, as well as tracing these indices back to the underlying wavelengths for each growing stage. This enables more targeted data collection and analysis in the future. 
\textcolor{red}{FeatureExplorer can also be applied to other sensor data (e.g., multispectral, LiDAR) that possess similar properties to hyperspectral indices (e.g. high dimensions, derived correlated features), to predict variables other than biomass.
Our system also can be adjusted to include different regression models since the underlying model will not intrinsically impact the feature exploration workflow.}



Future visual analytics research should investigate the dynamic generation of features based on raw input data, e.g. customized features based on different formulations of hyperspectral indices. 
\textcolor{red}{Also, one can improve the feature selection workflow by visually highlighting potential features in clusters that are ranked high importance (or low), for faster subgroup inclusion/exclusion.}
Feature selection in regression models for spatially and temporally heterogeneous data is also an open area for research.
Specifically, the geovisualization of feature importance for spatial regression methods has not been adequately addressed.
Finally, time series analysis can be incorporated to model temporally variable feature contributions, e.g. in a sequence of hyperspectral images with temporally variable wavelength reflectances at different plant growing stages.  

\acknowledgments{
The authors wish to thank Christina A. Stober and Hao Chen.
This work is funded by the Research Projects Agency-Energy, U.S. Department of Energy, under Award Number DE-AR0000593.
}

\bibliographystyle{abbrv-doi-narrow}

\bibliography{main}

\begin{thebibliography}{10}
\renewcommand*{\sfdefault}{PTSansNarrow-TLF}

\bibitem{alpaydin2009introduction}
E.~Alpaydin.
\newblock {\em Introduction to Machine Learning}.
\newblock MIT press, Cambridge, MA, USA, 2009.

\bibitem{Barlowe2008}
S.~Barlowe, T.~Zhang, Y.~Liu, J.~Yang, and D.~Jacobs.
\newblock {Multivariate visual explanation for high dimensional datasets}.
\newblock In {\em IEEE Symposium on Visual Analytics Science and Technology
  (VAST '08)}, pp. 147--154, Oct 2008. doi: \textsf{%
10\hspace{.1pt}\discretionary{.}{%
}{.}\hspace{.4pt}1109\discretionary{/}{%
}{/}VAST\hspace{.1pt}\discretionary{.}{%
}{.}\hspace{.4pt}2008\hspace{.1pt}\discretionary{.}{%
}{.}\hspace{.4pt}4677368}


\bibitem{brandao2015spectral}
Z.~N. Brand{\~a}o, V.~Sofiatti, J.~R. Bezerra, G.~B. Ferreira, J.~C. Medeiros,
  et~al.
\newblock Spectral reflectance for growth and yield assessment of irrigated
  cotton.
\newblock {\em Australian Journal of Crop Science}, 9(1):75--84, Jan 2015.

\bibitem{Chandrashekar2014}
G.~Chandrashekar and F.~Sahin.
\newblock A survey on feature selection methods.
\newblock {\em Computers {\&} Electrical Engineering}, 40(1):16--28, Jan 2014.
  doi: \textsf{%
10\hspace{.1pt}\discretionary{.}{%
}{.}\hspace{.4pt}1016\discretionary{/}{%
}{/}j\hspace{.1pt}\discretionary{.}{%
}{.}\hspace{.4pt}compeleceng\hspace{.1pt}\discretionary{.}{%
}{.}\hspace{.4pt}2013\hspace{.1pt}\discretionary{.}{%
}{.}\hspace{.4pt}11\hspace{.1pt}\discretionary{.}{%
}{.}\hspace{.4pt}024}


\bibitem{dasbeames}
S.~Das, D.~Cashman, R.~Chang, and A.~Endert.
\newblock {BEAMES}: Interactive multi-model steering, selection, and inspection
  for regression tasks.

\bibitem{Ding:2011:SBF:1943363.1943368}
J.~Ding, J.~Shi, and F.-X. Wu.
\newblock {SVM-RFE} based feature selection for tandem mass spectrum quality
  assessment.
\newblock {\em International Journal of Data Mining and Bioinformatics},
  5(1):73--88, Feb 2011. doi: \textsf{%
10\hspace{.1pt}\discretionary{.}{%
}{.}\hspace{.4pt}1504\discretionary{/}{%
}{/}IJDMB\hspace{.1pt}\discretionary{.}{%
}{.}\hspace{.4pt}2011\hspace{.1pt}\discretionary{.}{%
}{.}\hspace{.4pt}038578}


\bibitem{dingen2019regressionexplorer}
D.~{Dingen}, M.~{van't Veer}, P.~{Houthuizen}, E.~H.~J. {Mestrom}, E.~H. H.~M.
  {Korsten}, A.~R.~A. {Bouwman}, and J.~{van Wijk}.
\newblock {RegressionExplorer}: Interactive exploration of logistic regression
  models with subgroup analysis.
\newblock {\em IEEE Transactions on Visualization and Computer Graphics},
  25(1):246--255, Jan 2019. doi: \textsf{%
10\hspace{.1pt}\discretionary{.}{%
}{.}\hspace{.4pt}1109\discretionary{/}{%
}{/}TVCG\hspace{.1pt}\discretionary{.}{%
}{.}\hspace{.4pt}2018\hspace{.1pt}\discretionary{.}{%
}{.}\hspace{.4pt}2865043}


\bibitem{duan2005multiple}
K.-B. Duan, J.~C. {Rajapakse}, H.~Wang, and F.~{Azuaje}.
\newblock Multiple {SVM-RFE} for gene selection in cancer classification with
  expression data.
\newblock {\em IEEE Transactions on NanoBioscience}, 4(3):228--234, Sep 2005.
  doi: \textsf{%
10\hspace{.1pt}\discretionary{.}{%
}{.}\hspace{.4pt}1109\discretionary{/}{%
}{/}TNB\hspace{.1pt}\discretionary{.}{%
}{.}\hspace{.4pt}2005\hspace{.1pt}\discretionary{.}{%
}{.}\hspace{.4pt}853657}


\bibitem{elbahnasawy2018multi}
M.~Elbahnasawy, T.~Shamseldin, R.~Ravi, T.~Zhou, Y.-J. Lin, A.~Masjedi,
  E.~Flatt, M.~Crawford, and A.~Habib.
\newblock Multi-sensor integration onboard a {UAV}-based mobile mapping system
  for agricultural management.
\newblock In {\em IEEE International Geoscience and Remote Sensing Symposium
  (IGARSS '18)}, pp. 3412--3415, July 2018. doi: \textsf{%
10\hspace{.1pt}\discretionary{.}{%
}{.}\hspace{.4pt}1109\discretionary{/}{%
}{/}IGARSS\hspace{.1pt}\discretionary{.}{%
}{.}\hspace{.4pt}2018\hspace{.1pt}\discretionary{.}{%
}{.}\hspace{.4pt}8517370}


\bibitem{Elmqvist2008}
N.~Elmqvist, P.~Dragicevic, and J.-D. Fekete.
\newblock Rolling the dice: Multidimensional visual exploration using
  scatterplot matrix navigation.
\newblock {\em IEEE Transactions on Visualization and Computer Graphics},
  14(6):1539--1148, Nov 2008. doi: \textsf{%
10\hspace{.1pt}\discretionary{.}{%
}{.}\hspace{.4pt}1109\discretionary{/}{%
}{/}TVCG\hspace{.1pt}\discretionary{.}{%
}{.}\hspace{.4pt}2008\hspace{.1pt}\discretionary{.}{%
}{.}\hspace{.4pt}153}


\bibitem{Friendly2002}
M.~Friendly.
\newblock {Corrgrams: Exploratory displays for correlatigon matrices}.
\newblock {\em American Statistician}, 56(4):316--324, 2002. doi: \textsf{%
10\hspace{.1pt}\discretionary{.}{%
}{.}\hspace{.4pt}1198\discretionary{/}{%
}{/}000313002533}


\bibitem{gerik2003sorghum}
T.~Gerik, B.~Bean, and R.~Vanderlip.
\newblock Sorghum growth and development.
\newblock {\em Texas AgriLife Extension publication}, 2003.

\bibitem{Guo2009}
Z.~{Guo}, M.~O. {Ward}, and E.~A. {Rundensteiner}.
\newblock Model space visualization for multivariate linear trend discovery.
\newblock In {\em IEEE Symposium on Visual Analytics Science and Technology
  (VAST '09)}, pp. 75--82, Oct 2009. doi: \textsf{%
10\hspace{.1pt}\discretionary{.}{%
}{.}\hspace{.4pt}1109\discretionary{/}{%
}{/}VAST\hspace{.1pt}\discretionary{.}{%
}{.}\hspace{.4pt}2009\hspace{.1pt}\discretionary{.}{%
}{.}\hspace{.4pt}5333431}


\bibitem{Johansson2009}
S.~Johansson and J.~Johansson.
\newblock Interactive dimensionality reduction through user-defined
  combinations of quality metrics.
\newblock {\em IEEE Transactions on Visualization and Computer Graphics},
  15(6):993--1000, Nov 2009. doi: \textsf{%
10\hspace{.1pt}\discretionary{.}{%
}{.}\hspace{.4pt}1109\discretionary{/}{%
}{/}TVCG\hspace{.1pt}\discretionary{.}{%
}{.}\hspace{.4pt}2009\hspace{.1pt}\discretionary{.}{%
}{.}\hspace{.4pt}153}


\bibitem{Krause2014}
J.~Krause, A.~Perer, and E.~Bertini.
\newblock {INFUSE}: Interactive feature selection for predictive modeling of
  high dimensional data.
\newblock {\em IEEE Transactions on Visualization and Computer Graphics},
  20(12):1614--1623, Dec 2014. doi: \textsf{%
10\hspace{.1pt}\discretionary{.}{%
}{.}\hspace{.4pt}1109\discretionary{/}{%
}{/}TVCG\hspace{.1pt}\discretionary{.}{%
}{.}\hspace{.4pt}2014\hspace{.1pt}\discretionary{.}{%
}{.}\hspace{.4pt}2346482}


\bibitem{liang2015estimation}
L.~Liang, L.~Di, L.~Zhang, M.~Deng, Z.~Qin, S.~Zhao, and H.~Lin.
\newblock Estimation of crop {LAI} using hyperspectral vegetation indices and a
  hybrid inversion method.
\newblock {\em Remote Sensing of Environment}, 165:123--134, 2015. doi:
  \textsf{%
10\hspace{.1pt}\discretionary{.}{%
}{.}\hspace{.4pt}1016\discretionary{/}{%
}{/}j\hspace{.1pt}\discretionary{.}{%
}{.}\hspace{.4pt}rse\hspace{.1pt}\discretionary{.}{%
}{.}\hspace{.4pt}2015\hspace{.1pt}\discretionary{.}{%
}{.}\hspace{.4pt}04\hspace{.1pt}\discretionary{.}{%
}{.}\hspace{.4pt}032}


\bibitem{liu2011feature}
Q.~Liu, C.~Chen, Y.~Zhang, and Z.~Hu.
\newblock Feature selection for support vector machines with {RBF} kernel.
\newblock {\em Artificial Intelligence Review}, 36(2):99--115, Aug 2011. doi:
  \textsf{%
10\hspace{.1pt}\discretionary{.}{%
}{.}\hspace{.4pt}1007\discretionary{/}{%
}{/}s10462\discretionary{%
}{-}{-}011\discretionary{%
}{-}{-}9205\discretionary{%
}{-}{-}2}


\bibitem{masjedi2018sorghum}
A.~Masjedi, J.~Zhao, A.~M. Thompson, K.-W. Yang, J.~E. Flatt, M.~M. Crawford,
  D.~S. Ebert, M.~R. Tuinstra, G.~Hammer, and S.~Chapman.
\newblock Sorghum biomass prediction using {UAV}-based remote sensing data and
  crop model simulation.
\newblock In {\em IEEE International Geoscience and Remote Sensing Symposium
  (IGARSS '18)}, pp. 7719--7722, July 2018. doi: \textsf{%
10\hspace{.1pt}\discretionary{.}{%
}{.}\hspace{.4pt}1109\discretionary{/}{%
}{/}IGARSS\hspace{.1pt}\discretionary{.}{%
}{.}\hspace{.4pt}2018\hspace{.1pt}\discretionary{.}{%
}{.}\hspace{.4pt}8519034}


\bibitem{May2011}
T.~May, A.~Bannach, J.~Davey, T.~Ruppert, and J.~Kohlhammer.
\newblock Guiding feature subset selection with an interactive visualization.
\newblock In {\em IEEE Conference on Visual Analytics Science and Technology
  (VAST '11)}, pp. 111--120, Oct 2011. doi: \textsf{%
10\hspace{.1pt}\discretionary{.}{%
}{.}\hspace{.4pt}1109\discretionary{/}{%
}{/}VAST\hspace{.1pt}\discretionary{.}{%
}{.}\hspace{.4pt}2011\hspace{.1pt}\discretionary{.}{%
}{.}\hspace{.4pt}6102448}


\bibitem{Muhlbacher2013}
T.~Muhlbacher and H.~Piringer.
\newblock A partition-based framework for building and validating regression
  models.
\newblock {\em IEEE Transactions on Visualization and Computer Graphics},
  19(12):1962--1971, Dec 2013. doi: \textsf{%
10\hspace{.1pt}\discretionary{.}{%
}{.}\hspace{.4pt}1109\discretionary{/}{%
}{/}TVCG\hspace{.1pt}\discretionary{.}{%
}{.}\hspace{.4pt}2013\hspace{.1pt}\discretionary{.}{%
}{.}\hspace{.4pt}125}


\bibitem{Piringer2008}
H.~Piringer, W.~Berger, and H.~Hauser.
\newblock Quantifying and comparing features in high-dimensional datasets.
\newblock In {\em 12th International Conference Information Visualisation
  (InfoVis '08)}, pp. 240--245, July 2008. doi: \textsf{%
10\hspace{.1pt}\discretionary{.}{%
}{.}\hspace{.4pt}1109\discretionary{/}{%
}{/}IV\hspace{.1pt}\discretionary{.}{%
}{.}\hspace{.4pt}2008\hspace{.1pt}\discretionary{.}{%
}{.}\hspace{.4pt}17}


\bibitem{Piringer2010}
H.~Piringer, W.~Berger, and J.~Krasser.
\newblock {HyperMoVal}: Interactive visual validation of regression models for
  real-time simulation.
\newblock {\em Computer Graphics Forum}, 29(3):983--992, Aug 2010. doi:
  \textsf{%
10\hspace{.1pt}\discretionary{.}{%
}{.}\hspace{.4pt}1111\discretionary{/}{%
}{/}j\hspace{.1pt}\discretionary{.}{%
}{.}\hspace{.4pt}1467\discretionary{%
}{-}{-}8659\hspace{.1pt}\discretionary{.}{%
}{.}\hspace{.4pt}2009\hspace{.1pt}\discretionary{.}{%
}{.}\hspace{.4pt}01684\hspace{.1pt}\discretionary{.}{%
}{.}\hspace{.4pt}x}


\bibitem{scholkopf2002learning}
B.~Sch{\"o}lkopf and A.~J. Smola.
\newblock {\em Learning with Kernels: Support Vector Machines, Regularization,
  Optimization, and Beyond}.
\newblock MIT press, Cambridge, MA, USA, 2002.

\bibitem{JinwookSeo2004}
J.~Seo and B.~Shneiderman.
\newblock A rank-by-feature framework for unsupervised multidimensional data
  exploration using low dimensional projections.
\newblock In {\em IEEE Symposium on Information Visualization (InfoVis '04)},
  pp. 65--72, Oct 2004. doi: \textsf{%
10\hspace{.1pt}\discretionary{.}{%
}{.}\hspace{.4pt}1109\discretionary{/}{%
}{/}INFVIS\hspace{.1pt}\discretionary{.}{%
}{.}\hspace{.4pt}2004\hspace{.1pt}\discretionary{.}{%
}{.}\hspace{.4pt}3}


\bibitem{Brushing11}
C.~{Turkay}, P.~{Filzmoser}, and H.~{Hauser}.
\newblock Brushing dimensions - {A} dual visual analysis model for
  high-dimensional data.
\newblock {\em IEEE Transactions on Visualization and Computer Graphics},
  17(12):2591--2599, Dec 2011. doi: \textsf{%
10\hspace{.1pt}\discretionary{.}{%
}{.}\hspace{.4pt}1109\discretionary{/}{%
}{/}TVCG\hspace{.1pt}\discretionary{.}{%
}{.}\hspace{.4pt}2011\hspace{.1pt}\discretionary{.}{%
}{.}\hspace{.4pt}178}


\bibitem{Yang2003}
J.~Yang, W.~Peng, M.~O. Ward, and E.~A. Rundensteiner.
\newblock Interactive hierarchical dimension ordering, spacing and filtering
  for exploration of high dimensional datasets.
\newblock In {\em IEEE Symposium on Information Visualization (InfoVis '03)},
  pp. 105--112, Oct 2003. doi: \textsf{%
10\hspace{.1pt}\discretionary{.}{%
}{.}\hspace{.4pt}1109\discretionary{/}{%
}{/}INFVIS\hspace{.1pt}\discretionary{.}{%
}{.}\hspace{.4pt}2003\hspace{.1pt}\discretionary{.}{%
}{.}\hspace{.4pt}1249015}


\bibitem{zhang2017prediction}
Z.~Zhang, A.~Masjedi, J.~Zhao, and M.~M. Crawford.
\newblock Prediction of sorghum biomass based on image based features derived
  from time series of {UAV} images.
\newblock In {\em IEEE International Geoscience and Remote Sensing Symposium
  (IGARSS '17)}, pp. 6154--6157, July 2017. doi: \textsf{%
10\hspace{.1pt}\discretionary{.}{%
}{.}\hspace{.4pt}1109\discretionary{/}{%
}{/}IGARSS\hspace{.1pt}\discretionary{.}{%
}{.}\hspace{.4pt}2017\hspace{.1pt}\discretionary{.}{%
}{.}\hspace{.4pt}8128413}


\end{thebibliography}
\end{document}